\documentclass[showpacs]{revtex4}

\usepackage{amsmath}
\usepackage{amssymb}
\usepackage{graphicx}

\begin{document}

\title{Non-singular solutions to Einstein-Klein-Gordon equations
with a phantom scalar field}
\author{Vladimir Dzhunushaliev
\footnote{Senior Associate of the Abdus Salam ICTP}}
\email{dzhun@krsu.edu.kg}
\affiliation{ASC,
Department f{\" u}r Physik, Ludwig-Maximilians-Universit{\" a}t M{\" u}nchen,
Theresienstr. 37, D-80333, Munich, Germany \\
and \\
Department of Physics and Microelectronic
Engineering, Kyrgyz-Russian Slavic University, Bishkek, Kievskaya Str.
44, 720021, Kyrgyz Republic}

\author{Vladimir Folomeev}
\email{vfolomeev@mail.ru}
\affiliation{Institute of Physics of National Academy of Science
Kyrgyz Republic, 265 a, Chui Street, Bishkek, 720071,  Kyrgyz Republic}

\author{Ratbay Myrzakulov}
\email{cnlpmyra1954@yahoo.com, cnlpmyra@mail.ru}
\affiliation{Department of General and Theoretical Physics,
Eurasian National University, Astana, 010008, Kazakhstan}

\author{Douglas Singleton}
\email{dougs@csufresno.edu}
\affiliation{Physics Department, CSU Fresno, Fresno, CA 93740-8031}

\begin{abstract}
It is shown that the 4D Einstein-Klein-Gordon equations with a phantom scalar field
(a scalar field with a negative sign in front of the kinetic energy term of its
Lagrange density) has non-singular, spherically symmetry solutions. These solutions
have a combination of features found in other spherically symmetric gravity plus field solutions.
A stability analysis on these solutions indicates they are unstable.
\end{abstract}


\maketitle

\section{Introduction}

When one considers the system of gravity plus some field(s) (e.g. scalar field, gauge fields)
there are a range of spherically symmetric solutions. When these extra fields are non-interacting,
complex, scalar fields one finds the boson star solutions \cite{Fein, Kaup, Ruff} which
are prevented from collapsing because of the Heisenberg uncertainty principle. Allowing
the complex scalar field to have a self interaction of the $\lambda \phi ^4$ type lead
to boson stars with masses comparable to the Chandrasekhar mass \cite{Colpi,Miel} in contrast
to the much smaller mass of the non-interacting boson star solutions of \cite{Fein, Kaup, Ruff}.

If one considers a real scalar field rather than a complex scalar then there are no non-singular, static
solutions \cite{Kodama:1978dw, Baekler:1987jb, schmolt, jetzer,Sushkov:2002ef}. However it was also shown in \cite{Kodama:1978dw}
that if one considers ghost fields (a real scalar field with a negative sign for the kinetic {\it and}
potential parts of the scalar field Lagrangian density) one has non-singular, static solutions, but having a
non-trivial topology. These ghost field solutions have a wormhole structure. This might have
been expected since ghost fields may have a negative mass-energy density,
and it is just such a form of matter that is needed in order to  support wormholes \cite{morris}.
A similar situation occurs when one considers wormhole-like solutions with a non-minimally coupled
scalar field having a non-trivial topology~\cite{Sushkov:2002ef}. A final related system occurs
with a phantom Born-Infeld field forming compact objects (``gravastars") \cite{Bilic:2005sn}.

In this work we investigate spherically symmetric solutions to gravity plus a real scalar field. In contrast
to previous work we consider a phantom field \cite{caldwell} i.e. a real scalar field with a negative sign in
front of {\it only} the kinetic energy part of the Lagrangian density. Phantom fields were originally
proposed as an extreme form of dark energy which could explain the accelerated expansion of the
Universe \cite{perlmutter}. Phantom fields can violate the weak energy condition (WEC) and appear to be
quantum mechanically unstable. (The WEC requires that the mass-energy density, $\varepsilon$, and
pressure, $p$, of the field/fluid satisfy $\varepsilon + p \ge 0$ or $w \equiv p/\varepsilon \ge -1$).
Any field or fluid with $w <-1/3$ will lead to gravitationally repulsion and is called dark energy.
Phantom energy with $w <-1$ is an extreme form of dark energy.
A field or fluid with $w > -1/3$ leads to gravitational attraction.
Nevertheless, as pointed out in \cite{caldwell} the supernova data seem
to favor an extreme form of dark energy like phantom energy.
The original proposal for phantom energy can be found in \cite{bronn, ellis}
(see also \cite{Feinstein:2003iz} and references therein for related work in this direction). There are also the so-called
``quintom'' models of dark energy~\cite{Cai:2007qw} with an
equation of state that crosses the cosmological constant boundary and, correspondingly,
violates the WEC.

Phantom energy can lead to gravitational repulsion and therefore accelerated expansion of the Universe. This
repulsion can provide a mechanism for the existence of static, regular solutions -- the gravitational repulsion
of the phantom field can be balanced by the gravitational attraction of normal matter. This is the physical
mechanism behind the Bartnik-McKinnion solution \cite{bartnik} to the Einstein-Yang-Mills system --
the repulsion coming for the self-interaction of the Yang-Mills fields balances the usual gravitational
attraction and stabilizes the configuration. As with the Bartnik-McKinnion solution the phantom field
solution presented here is regular everywhere and asymptotically goes to Minkowski spacetime. Unlike the ghost
field, kink solution of \cite{Kodama:1978dw}, with its non-trivial, wormhole topology, the present phantom field
solution (like the Bartnik-McKinnion solution) represents a spherically symmetric, localized field configuration,
and has a trivial topology.

One unusual feature of the phantom field solutions examined here is that it has an overall
negative mass. This is in contrast to the ghost field solution of \cite{Kodama:1978dw}
with its positive mass. Both phantom fields and ghost fields face difficulties
when one tries to quantize them. Nevertheless the data on the accelerated expansion rate of the Universe
favors something like phantom energy \cite{caldwell} and so one should study the consequences of such
unusual form of matter-energy. As far as we know phantom energy has been studied in a cosmological context where
the phantom field varies only with time but does not vary spatially. In the present paper our phantom
field varies spatially. Also strictly speaking our scalar field is only a phantom field (i.e. a field
with $w <-1$) in some spatial regions, while in other regions it behaves as a gravitationally
attractive field with $w > -1/3$. Physically this makes sense since a field which is phantom everywhere
in space could not form some static spherically symmetric configuration but would disperse.

There have been other works which have studied spherically symmetric solutions
with phantom or ghost fields \cite{picon, lobo,sushkov}. In \cite{bronn2} black
hole solutions in the presence of a phantom field were investigated.
In \cite{Babichev} the process of accretion of phantom energy onto a black hole
was considered.
In \cite{Shat} a study was undertaken of spherically symmetric solutions for gravity
plus some matter/fluid which violated weak energy condition. In this paper we
replace the fluid by a scalar field. In regard to the negative mass of our solution
there are other known negative mass solutions such as the negative mass
black holes of \cite{mann}. However as with the ghost field kink solutions
of \cite{Kodama:1978dw} these negative mass black hole solutions have non-trivial
topology.

\section{Equations and solutions}

We chose our spherically symmetric metric to have the form \cite{Kodama:1978dw}
\begin{equation}
\label{metric_sch}
ds^2=e^{\nu(t,r)}dt^2-e^{\lambda(t,r)}dr^2-r^2(d\theta^2+\sin^2\theta d\phi^2).
\end{equation}
For the matter Lagrangian we have one real phantom scalar field with the Lagrangian
\begin{equation}
\label{lagrangian}
  L =-\frac{R}{16\pi G}-
      \frac{1}{2}\partial_\mu \varphi \partial^\mu
        \varphi -V(\varphi)~.
\end{equation}
For the potential we take
\begin{equation}
\label{pot}
V(\varphi)=-\frac{1}{2}m^2\varphi^2+\frac{\kappa}{4}\varphi^4,
\end{equation}
where $m$ is the mass of the scalar field and $\kappa$ is the coupling constant.
This is the usual ``Mexican hat" symmetry breaking potential. Note that in our terminology
phantom field means a negative sign in front of the kinetic energy term only. By ghost
field we mean a negative sign in front of both the kinetic and potential terms of the
scalar field Lagrangian. With this definition it is phantom fields (not ghost fields) which
can lead to $w < -1$ and accelerated expansion of the Universe.  In what follows
we set $16\pi G=1$.

The energy-momentum tensor follows from \eqref{lagrangian}
\begin{equation}
\label{EMT}
  T_{\mu\nu}=-\partial_{\mu}\varphi\partial_{\nu}\varphi-g_{\mu\nu}
  \left[-
      \frac{1}{2}\partial_\alpha \varphi \partial^\alpha
        \varphi -V(\varphi)\right].
\end{equation}
The corresponding Einstein equations are
\begin{equation}
\label{Ein_gen}
R_{\mu\nu}=\frac{1}{2}\left[-\partial_\mu\varphi\partial_\nu\varphi-g_{\mu\nu}V(\varphi)\right]
\end{equation}
Note we are using the conventions of \cite{landau} where the Ricci tensor is defined
via the contraction of the first and third index of the Riemann tensor
$R_{\mu \nu } \equiv R^\alpha _{\mu \alpha \nu}$. This is the reason for the overall
$+$ sign on the right hand side of \eqref{Ein_gen} (see also \cite{clayton}).
From \eqref{Ein_gen} it follows
\begin{eqnarray}
\label{Ein_comp}
R_{01}&=&\frac{1}{r}\dot{\lambda}=-\frac{1}{2}\dot{\varphi} \varphi^\prime,
\nonumber\\
e^{-\nu}R_{00}+e^{-\lambda}R_{11}&=&\frac{1}{r}e^{-\lambda}\left(\nu^\prime+\lambda^\prime\right)=
-\frac{1}{2}e^{-\nu}\dot{\varphi}^2-\frac{1}{2}e^{-\lambda}\varphi^{\prime 2},\\
R_{22}&=&-\frac{1}{2}r e^{-\lambda}\left(\nu^\prime-\lambda^\prime\right)+1-e^{-\lambda}=\frac{1}{2}r^2 V(\varphi).\nonumber
\end{eqnarray}
The wave equation for the scalar field $\varphi$ follows from
variation of the Lagrangian \eqref{lagrangian}
\begin{equation}
\label{sfe}
e^{-\nu}\ddot{\varphi}-\frac{1}{2}e^{-\nu}\left({\dot \nu}-{\dot \lambda}\right)\dot{\varphi}-e^{-\lambda}\varphi^{\prime\prime}-
e^{-\lambda}\left[\frac{2}{r}+\frac{1}{2}\left(\nu^\prime-\lambda^\prime\right)\right]\varphi^\prime+\frac{d V}{d\varphi}=0.
\end{equation}
In the above the ansatz functions are $t$ and $r$ dependent. Differentiation with respect to $t$ ($r$) is denoted by
a $dot$ ($prime$).

To investigate the static problem we drop the time dependence and express the metric functions
$\nu, \lambda$ via new metric functions $C(r), M(r)$
\begin{equation}
\label{conn}
e^{\nu}=C(r)\left[1-\frac{2 M(r)}{r}\right], \qquad
e^{\lambda}=\left[1-\frac{2 M(r)}{r}\right]^{-1}.
\end{equation}
(In the next section when we investigate the stability of our solutions we will again consider time
dependent ansatz functions). In order to perform the numerical analysis we also introduce new dimensionless
variables
$$x=m r,\quad \tilde{M}=2 m M,\quad
\Lambda=\frac{\kappa}{m^2}.$$
Using these functions and variables the Einstein equations \eqref{Ein_comp} become
\begin{equation}
\label{Ein_1}
\frac{1}{x}\frac{C^\prime}{C}\left(1-\frac{\tilde{M}}{x}\right)-\frac{\tilde{M}^\prime}{x^2}=
\frac{1}{4}\left[-\left(1-\frac{\tilde{M}}{x}\right)\varphi^{\prime 2}+\varphi^2-\frac{\Lambda}{2}
\varphi^4\right]
\end{equation}
and the equation for $\tilde{M}$ becomes
\begin{equation}
\label{Ein_2}
\tilde{M}^\prime=-\frac{1}{4}x^2\left[\left(1-\frac{\tilde{M}}{x}\right)\varphi^{\prime 2}+\varphi^2-\frac{\Lambda}{2}
\varphi^4\right]~.
\end{equation}
Combining \eqref{Ein_1} and \eqref{Ein_2} we find
\begin{equation}
\label{Ein_3}
\frac{C^\prime}{C}=-\frac{1}{2}x\varphi^{\prime 2}.
\end{equation}
Also the scalar field equation \eqref{sfe} takes the form
\begin{equation}
\label{sfe_n}
\varphi^{\prime\prime}+\left\{\frac{2}{x}\left[1-\frac{1}{2}\left(\frac{\tilde{M}^\prime-\tilde{M}/x}
{1-\tilde{M}/x}\right)\right]-\frac{1}{4}x\varphi^{\prime 2}\right\}\varphi^{\prime}=
\left(1-\frac{\tilde{M}}{x}\right)^{-1}\varphi \left(1-\Lambda\varphi^2\right).
\end{equation}
Thus we only needed to solve for $M(x), \varphi (x)$ via \eqref{Ein_2} \eqref{sfe_n} since
from \eqref{Ein_3} $C(x)$ was determined by $\varphi(x)$.

We solved equations \eqref{Ein_2} and \eqref{sfe_n} numerically
using  the following boundary conditions:
\begin{equation}
\varphi(0)= const, \qquad \varphi^\prime(0)=0, \qquad \tilde{M}(0) =0.
\end{equation}
That is we demand regularity at the origin. Next choosing some $\Lambda$ we found
finite energy, non-singular solutions by numerically solving the system of equations
\eqref{Ein_2} and \eqref{sfe_n}. For a given $\Lambda$ this amounted to finding the correct
initial value of the scalar field $\varphi(0)$ which gave the desired well behaved
solution. The problem reduces effectively to an eigenvalue problem since for a given
$\Lambda$ one had to search for the proper ``eigenvalue" $\varphi(0)$ in order to
get a well behaved solution (regular, asymptotically flat) for the metric and scalar field
ansatz functions. The flatness of the spacetime at infinity was provided by choosing the proper initial
value for $C(0)$ in equation \eqref{Ein_3} i.e. we could chose $C(0)$ such that
$C(\infty)\rightarrow 1$.

\begin{figure}[t]
\begin{minipage}[t]{.49\linewidth}
\begin{center}
  \includegraphics[width=9cm]{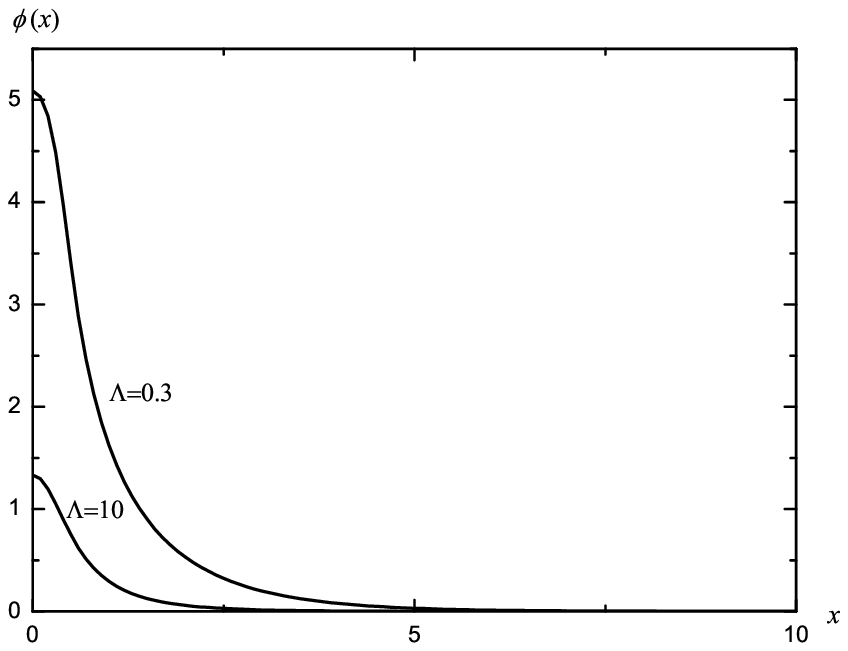}
\vspace{-1cm}
 \caption{The scalar field $\varphi$ for two different values of $\Lambda$.}
\label{phi}
\end{center}
\end{minipage}\hfill
\begin{minipage}[t]{.49\linewidth}
  \begin{center}
  \includegraphics[width=9cm]{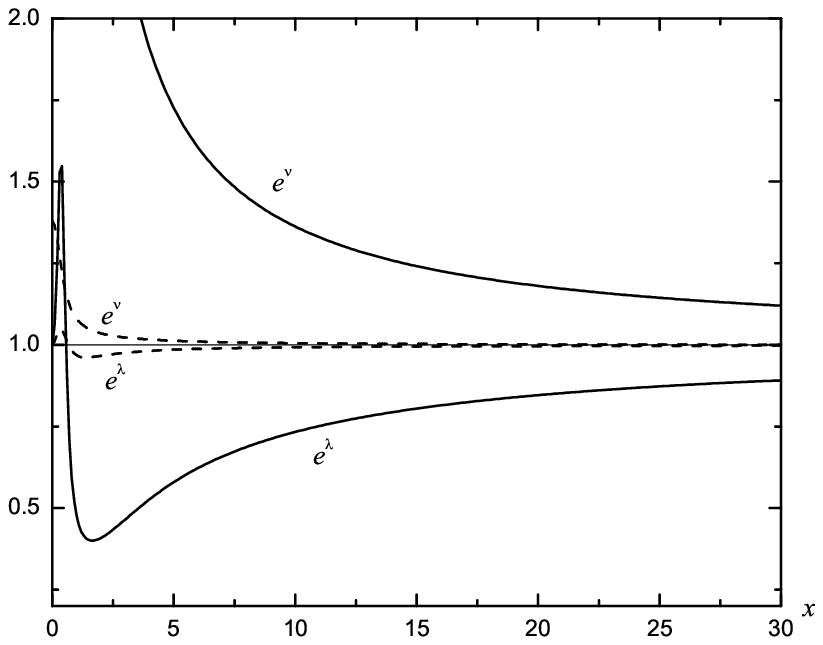}
\vspace{-1.5cm}
 \caption{The metric functions $e^{\nu(x)}$ and $e^{\lambda(x)}$ when $\Lambda=0.3$ (solid line) and $\Lambda=10$ (dashed line).}
\label{metr}
\end{center}
\end{minipage}\hfill
\end{figure}

In figure \ref{phi} we show the graphs of the scalar field for two values of $\Lambda$. Generically
the solutions for the scalar field looked similar -- starting at some non--zero value at $x=0$ and
decaying to zero as $x \rightarrow \infty$ i.e. $\varphi (\infty) = 0$. For a normal
scalar field this asymptotic value would be strange since for the ``Mexican hat" potential
this is a local maximum. Keeping in mind the unusual character of the phantom
field it is not unexpected that the field should asymptotically go to
a maximum rather than a minimum. The behavior of the metric functions $e^{\nu(x)}$ and
$e^{\lambda(x)}$ for these values of $\Lambda$ is shown in figure \ref{metr}. Again the
behavior of the metric functions is generally similar for different values of $\Lambda$ --
starting at some constant value at $x=0$ they asymptotically go to $1$ without ever becoming
$0$ or negative. Thus the metric is non-singular and without horizons.

\begin{figure}[t]
\begin{minipage}[t]{.49\linewidth}
\begin{center}
  \includegraphics[width=9cm]{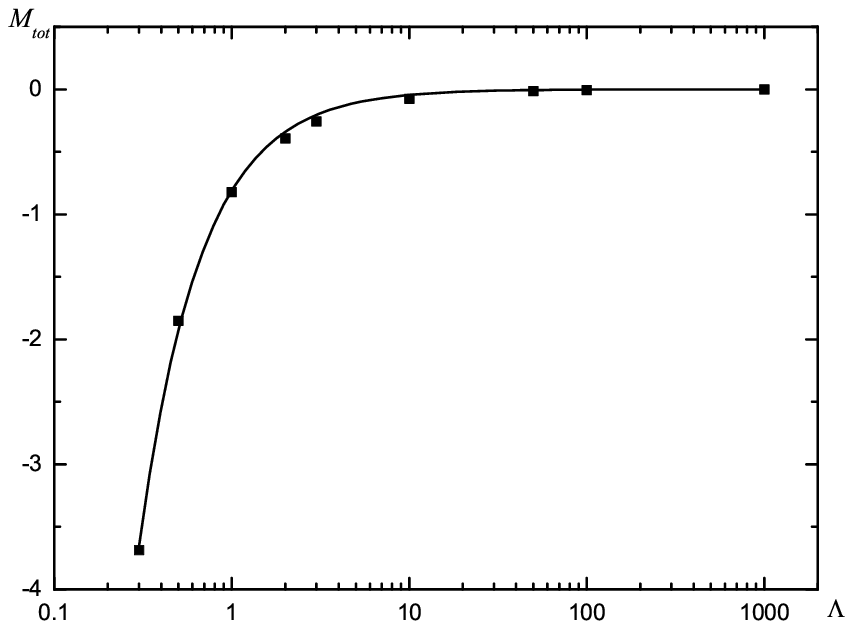}
 \vspace{-1cm}
 \caption{The dependence of dimensionless mass $\tilde{M} (\infty)$ on $\Lambda$.
 Here small squares show calculated values of the mass and the line
 is an approximation function with
equation $\approx const/\Lambda^{1.25}$.}
\label{M_lambda}
\end{center}
\end{minipage}\hfill
\begin{minipage}[t]{.49\linewidth}
  \begin{center}
  \includegraphics[width=9cm]{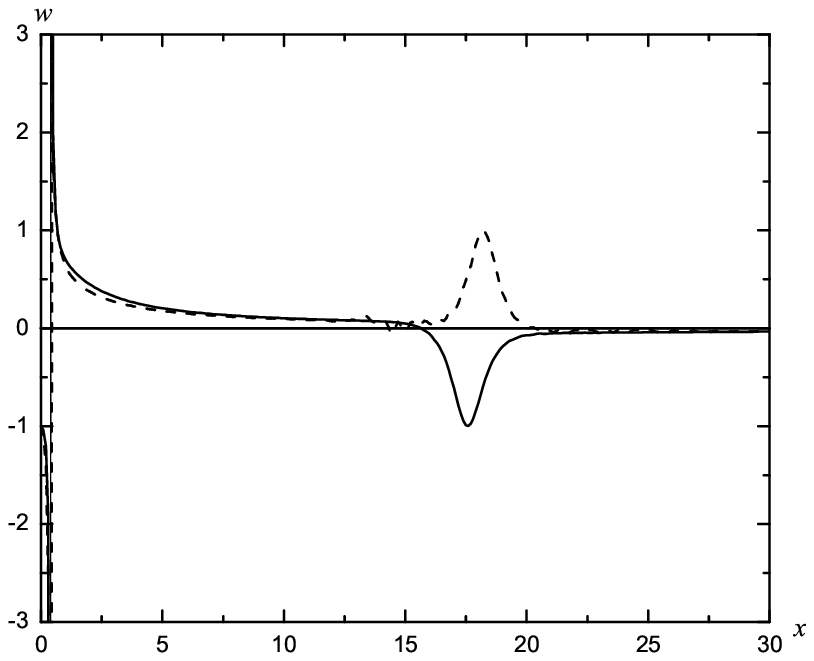}
  \vspace{-1.5cm}
  \caption{The equation of state $w(x)$ when $\Lambda=0.3$ (solid line) and $\Lambda=10$ (dashed line).}
  \label{eqs_f}
  \end{center}
\end{minipage}\hfill
\end{figure}

The mass of the solutions was obtained from the asymptotic values of $\tilde{M} (\infty)$
from \eqref{Ein_2} . The mass scale of the dimensionless quantity $\tilde{M} (\infty)$ is set
by the mass $m$ from the potential $V(\varphi)$. The dependence of the $\varphi (0)$ and $\tilde{M} (\infty)$ is shown in
Table I and also in figure \ref{M_lambda}. From the figure and table one can see that the
total mass goes to zero as $\Lambda\rightarrow \infty$. Also $\varphi (0)$ goes to
zero as $\Lambda\rightarrow \infty$. $\tilde{M} (\infty)$ approaches zero from below while
$\varphi (0)$ approaches zero from above.

\begin{table}[h]
\label{tab}
\caption{The dependence of $\phi(0)$ and $\tilde{M} (\infty)$ on $\Lambda$.}
\begin{center}
{\begin{tabular}{ccccc}
\multicolumn{3}{c}{} \\[10pt]\hline
$\varphi(0) $  & $\Lambda$ & $\tilde{M} (\infty)$ \\
\hline 7.1312051565&0.1&-23.8587\\
\hline 5.09014118325&0.3&-3.68446\\
\hline 4.35754&0.5&-1.85183\\
\hline 3.478853&1&-0.821492\\
\hline 2.697266&2&-0.391964\\
\hline 2.288345&3&-0.256198\\
\hline 1.3326834&10&-0.0755891\\
\hline 0.6098228&50&-0.0152216\\
\hline 0.4324704&100&-0.00756343\\
\hline 0.1371201&1000&-0.000753043\\
\hline
\end{tabular}}
\end{center}
\end{table}

\subsection{Asymptotic Behavior}

We now examine the asymptotic behavior of the solutions in order to get some analytic expressions for
the solutions for large $x$ and to compare to the numerical solutions. For this purpose, we write the asymptotic
forms of the functions as:
\begin{equation}
\label{asymptotic}
    \varphi\approx \delta \varphi, \quad C\approx C_{\infty}+\delta C, \quad
    \tilde{M} \approx \tilde{M}_{\infty}+\delta \tilde{M},
\end{equation}
where $\delta \phi, \delta C, \delta \tilde{M}\ll 1$. Then the scalar field
equation \eqref{sfe_n} becomes
\begin{equation}
\delta \varphi^{\prime \prime}+\frac{2}{x}\delta \varphi^\prime=\delta\varphi.
\end{equation}
We neglect terms of order $(\delta \varphi)^2$ , $1/x^2$ and higher. In order to neglect the term
$\Lambda (\delta \varphi)^2$ we need to additionally require that $\Lambda$ is not too large.
Although note that as $\Lambda$ becomes large $(\delta \varphi)$ goes more rapidly to
zero. This last equation has the asymptotic solution:
\begin{equation}
\delta \varphi \approx C_{\varphi} \frac{\exp{(-x)}}{x}.
\end{equation}
Inserting this solution into \eqref{Ein_2} gives the following equation and solution for $\delta \tilde{M}$:
\begin{equation}
\delta \tilde{M}^\prime\approx-\frac{1}{2}C_{\varphi}^2 \exp{(-2 x)}\quad \Rightarrow \quad
\delta \tilde{M}\approx \frac{1}{4} C_{\varphi}^2 \exp{(-2 x)}.
\end{equation}
Using these results in \eqref{Ein_3} we have the following equation and solution for $\delta C$:
\begin{equation}
\delta C^\prime\approx-\frac{1}{2}C_\infty C_{\varphi}^2 \frac{\exp{(-2 x)}}{x} \quad \Rightarrow \quad
\delta C \approx -\frac{1}{2} C_\infty C_{\varphi}^2 \,\text{Ei}(-2 x),
\end{equation}
where $\text{Ei}$ is the exponential integral function which goes to 0 from below as $x\rightarrow \infty$.
This shows that all perturbations ($\delta \phi, \delta C, \delta \tilde{M}$) tend asymptotically to zero
and that the ansatz functions as given in \eqref{asymptotic} approach the values given by the numerical
solution.

\subsection{Behavior of the $w$ parameter}

We now examine the behavior with respect to $x$ of the effective equation of state $w(x)=p(x)/\varepsilon(x)$ of
the scalar field ($\varepsilon(x)$ and $p(x)$ are respectively the energy density and pressure
of the scalar field). From~\eqref{metric_sch},\eqref{EMT} and \eqref{conn} we have:
\begin{eqnarray}
T^0_0&=&\varepsilon(x)=-\frac{1}{2}\left(1-\frac{\tilde{M}}{x}\right)\varphi^{\prime 2}+V(\varphi),\\
T^1_1&=&-p(x)=\frac{1}{2}\left(1-\frac{\tilde{M}}{x}\right)\varphi^{\prime 2}+V(\varphi).
\end{eqnarray}
The corresponding effective equation of state for the scalar field is:
\begin{equation}
\label{eqs}
w(x)=\frac{p(x)}{\varepsilon(x)}=\frac{-\frac{1}{2}\left(1-\frac{\tilde{M}}{x}\right)\varphi^{\prime 2}-V(\varphi)}
{-\frac{1}{2}\left(1-\frac{\tilde{M}}{x}\right)\varphi^{\prime 2}+V(\varphi)}.
\end{equation}
Using the numerical solution obtained earlier, figure \ref{eqs_f} shows the behavior of $w(x)$ for the
case when $\Lambda=0.3$ (solid line) and $\Lambda=10$ (dashed line). From the figure one sees
that there is some point $x=x_*$ in which the denominator of \eqref{eqs} goes to zero
(in the case under consideration this occurs around $x_*\approx 0.4$).
As $x\rightarrow x_*$ from the left $w\rightarrow -\infty$ and as $x\rightarrow x_*$ from
the right $w\rightarrow +\infty$. Asymptotically $w(x)$ tends to zero.
In the range $0<x<x_*$ we have $w\leq -1$ ($w=-1$ at $x=0$) and the WEC is violated. Thus in this
region our scalar field acts as phantom dark energy \cite{Amen}. In the range $x_*<x<\infty$ we have mostly
gravitationally attractive matter with $w>-1/3$. For $\Lambda =0.3$ there is some region around $x \approx 17$
where $w$ briefly dips down and just reaches $w=-1$. In this region the field is dark energy (but not phantom
energy) and is gravitationally repulsive. However for both $\Lambda =0.3$ and $\Lambda =10$ the field
is all or mostly gravitationally attractive. This provides a physical basis for the formation of
these solutions: the gravitational repulsion coming from the regions with phantom energy ($w < -1$) or
dark energy ($-1 < w < -1/3$) is balanced by the gravitational attraction coming from the regions
where the field has $w >-1/3$. This is reminiscent of the Bartnik-McKinnion solution which owes its
existence to the interplay with the repulsion of the Yang-Mills field against the attraction from
gravity. In the present case the repulsion of the regions where the scalar field has $w <-1/3$ (and
especially regions where $w <-1$) is balanced by the attraction of the regions where $w >-1/3$.

There have been studies \cite{Kamen} where the equation of state of some fluid or field, as given by $w$, varies with time.
In these studies there is some point in time where scalar field makes a transition and begins to violate the WEC
and becomes a phantom field. In the present work our equation of state, as given by $w$, is spatially varying and it
is this feature which gives rise to the static, spherically symmetric solutions.

A similar result can occur even for a regular, real, scalar coupled to gravity.  In \cite{Moffat:2007ch} such
a system was studied and it was found that there are regions where the weak energy condition is violated.
This paper also found non-singular spherically symmetric solutions.

\section{Stability analysis}

We now study the dynamical stability of the above solutions against linear perturbations.
We perturb the solutions of the system \eqref{Ein_comp} by expanding the metric functions and scalar
field function to first order as follows \cite{jetzer} \cite{clayton} \cite{torii}
\begin{eqnarray}
\nu(t,x)&=&\nu_0(x)+\nu_1(x) \cos(\omega t),\nonumber\\
\lambda(t,x)&=&\lambda_0(x)+\lambda_1(x) \cos(\omega t),\nonumber\\
\varphi(t,x)&=&\varphi_0(x)+\varphi_1(x) \cos(\omega t)/x.\nonumber
\end{eqnarray}
The index $0$ indicates the static background solutions of equations \eqref{Ein_2}-\eqref{sfe_n}. (Expressions
for $\nu_0(x), \,\lambda_0(x)$ can be obtained from \eqref{conn}.) Then the first-order perturbation
equations following from \eqref{Ein_comp} and \eqref{sfe} are
\begin{eqnarray}
\label{metr_pert}
\lambda_1&=&-\frac{1}{2}\varphi_0^\prime\varphi_1,\\
\nu_1^\prime&=&\frac{1}{2}\varphi_0^{\prime\prime}\varphi_1-\frac{1}{2}\varphi_0^\prime\varphi_1^\prime+\frac{1}{x}\varphi_0^\prime\varphi_1,
\end{eqnarray}
and the equation for $\varphi_1$ is
\begin{equation}
\label{sf_pert}
\varphi_1^{\prime\prime}+\frac{1}{2}\left(\nu_0^\prime-\lambda_0^\prime\right)\varphi_1^\prime-V_0(x)\varphi_1+
\omega^2 e^{\lambda_0-\nu_0} \varphi_1=0
\end{equation}
with the potential
\begin{equation}
\label{pot_pert}
V_0(x)=\frac{1}{2x}\left(\nu_0^\prime-\lambda_0^\prime\right)-\frac{1}{2}\varphi_0^{\prime\prime}\varphi_0^{\prime}x-
\frac{1}{2}\varphi_0^{\prime 2}-e^{\lambda_0}\left(1-3\Lambda\varphi_0^2\right).
\end{equation}

Introducing the new independent variable $\rho$
$$
\frac{d\rho}{d x}=e^{(\lambda_0-\nu_0)/2}
$$
we can rewrite equation \eqref{sf_pert} in a Schr\"{o}dinger-like form
$$
-\frac{d^2\varphi_1}{d\rho^2}+V[x(\rho)]\varphi_1=\omega^2\varphi_1,
$$
where $V[x(\rho)]=e^{\nu_0-\lambda_0}V_0(x)$. If there is a negative eigenvalue $\omega^2$ then the
solution will be unstable since then $\varphi_1 \sim e^{i \omega t}$ will grow exponentially.
To this end we examine the asymptotic behavior of the potential $V[x(\rho)]$. Using the asymptotic expressions
for the metric and scalar field ansatz functions from the previous section one finds
$$
V[x(\rho)]\rightarrow -e^{\nu_0}=-1 \quad \text{as} \quad x\rightarrow \infty.
$$

On the other hand, as $x\rightarrow 0$ we can expand the solutions into series in the following form
\begin{eqnarray}
\varphi_0&\sim& \varphi(0)+\frac{1}{6}\varphi(0)\left[1-\Lambda \varphi(0)^2\right]x^2,\nonumber\\
\tilde{M}_0&\sim& -\frac{1}{12}\left[ \varphi(0)^2-\frac{\Lambda}{2} \varphi(0)^4\right]x^3, \nonumber\\
C_0&\sim&C(0) \exp\left\{-\left[\frac{1}{3\sqrt{2}}\varphi(0)\left(1-\Lambda \varphi(0)^2\right)\right]^2\frac{x^4}{4}\right\}, \nonumber
\end{eqnarray}
whence the expressions for $\nu_0$ and $\lambda_0$ can be found (see equation \eqref{conn}). Then, using \eqref{pot_pert},
one finds that as $x\rightarrow 0$
$$
V[x(\rho)]\sim \alpha+\beta x^2,
$$
where
$$
\alpha = \gamma - C(0)\left[ 1 -3 \Lambda \varphi (0)^2-\gamma\right], \quad
\beta = C(0) \left\{\Lambda \varphi (0)^2 \left[1-\Lambda \varphi (0)^2 \right] - \gamma \right\},
$$
with
$$
\gamma=\frac{1}{12}\left[1-\frac{\Lambda}{2}\varphi(0)^2\right]\varphi(0)^2.
$$
Depending on $\Lambda$ and $\varphi (0)$ the coefficients $\alpha$ and $\beta$ can take values such 
that near $x=0$ one has a positive definite potential well or not. However 
since $V[x(\rho)] \rightarrow -1$ as $x \rightarrow \infty$ one always has some negative 
$\omega^2$ coming from continuum states. Because of these negative eigenvalues the solution is unstable.

The presence of unstable solutions is not surprising. It is well-known \cite{derrick} that in 4D there are no stable soliton solutions
to the Klein-Gordon equation with the ``Mexican hat" potential (or any other type of potential). In \cite{jetzer}
it was found that adding gravity did not give stable solutions. Here we have added the additional ingredient
of a phantom field but as the above stability analysis shows this also does not lead to stable
solutions. 

\section{Conclusions}

By studying the system of 4D gravity plus a phantom scalar field (a real scalar field with a
negative sign in front of the kinetic energy term in the Lagrangian density) we have
found finite energy, regular solutions which have trivial topology. The solutions share
some common features with previous solutions to systems of 4D gravity plus some
field(s). Like the Bartnik-McKinnion solutions our solutions
have finite energy, are regular everywhere, and have no horizons. The physical mechanism
behind the present solutions is similar to that of the Bartnik-McKinnion solutions: the existence
of the Bartnik-McKinnion solutions comes from the interplay of the repulsion of the
Yang-Mills fields against the attraction of gravity. The present solutions arise from
the balancing of regions where the scalar field a phantom/dark equation of state (and is
therefore gravitationally repulsive), against regions where the scalar field has the equation
of state for ordinary matter-energy (and is therefore gravitationally attractive).

The present solutions had some different features compared to the closely related
ghost field, kink solutions of \cite{Kodama:1978dw} (for these solutions there was a
negative sign in front of {\it both} the kinetic and potential terms in the
scalar field Lagrangian density). These kink solutions had a non-trivial wormhole
topology which resulted in their stability. In contrast the present solutions had a
trivial topology, an everywhere regular metric and no horizons. However the linear stability analysis
of the present solutions showed that they are unstable, as are the solutions of a
regular scalar field plus 4D gravity and the higher Bartnik-McKinnion solutions (those with
$k >2$). One possibility toward finding stable phantom field solutions might be to
alter the form of the potential $V(\varphi)$. Note that asymptotically
$\varphi \rightarrow 0$ which is an {\it unstable} equilibrium point for a
normal scalar field. This behavior is due to the reversed character of a scalar
field with a negative sign in front of kinetic energy term -- it should go toward maxima
of the potential. Unfortunately for the ``Mexican hat" potential $\varphi =0$ is only a local
maximum; the global maxima is at $\varphi = \pm \infty$. This suggests trying some potential
which has a global maxima at some finite $\varphi$ and with $V(\varphi)$ finite at this point.

The mass of our solutions turned out to be negative. For small $\Lambda$ the magnitude of the mass
was large and decreased to zero from below as $\Lambda \rightarrow \infty$ (see
figure \ref{M_lambda} and table I). Given the unusual nature of the
scalar field this could have been expected, and is similar to negative mass black hole
solutions \cite{mann}. Despite the instability and negative mass features of the
present solutions there is nevertheless a physical motivation for studying this
system since experimental evidence \cite{caldwell} favors the existence of something
like phantom energy.

\section{Acknowlegdment} V.F. was supported by a CSM International Activities Grant during the
completion of this work. V.F would also like to thank the CSU Fresno Physics Department for
their hospitality during his visit. V.D. is grateful to the Alexander von Humboldt Foundation
for financial support and V. Mukhanov for invitation to Universit\"at M\"unich for research.

\end{document}